# Performance evaluation of the cavities on nucleate boiling at microscale level

Yu-Tong Mu [a], Li Chen [a, b], Qin-Jun Kang [b] and Wen-Quan Tao [a, *]

a. Key Laboratory of Thermo-Fluid Engineering and Science of MOE, School of Energy and Power Engineering, Xi'an Jiaotong University, Xi'an, Shaanxi, 710049, China

b. Computational Earth Science Group (EES-16), Los Alamos National Laboratory, Los Alamos, NM, USA

*wqtao@mail.xjtu.edu.cn

## Abstract

Nucleate boiling heat transfer (NBHT) from enhanced structures is an effective way to dissipate high heat flux. In the present study, the cavities behaviours for nucleation on roughened surface are numerically studied on the entire ebullition cycle based on a phase-change lattice Boltzmann method without introducing any artificial disturbances. The adopted model is firstly validated with the Laplace law and the two phase coexistence curve, and then applied to investigate the effects of cavity structure on NBHT. The bubble departure diameter, departure frequency and total boiling heat flux of an ebullition cycle are also explored. It is demonstrated that the cavity widths and the cavity grooves show significant influence on the features of NBHT. Cavity with circular groove in the present research shows the best performance for NBHT in terms of the averaged heat flux and bubble release frequency. When a specific cavity is combined with other different cavities on roughened surfaces its nucleation process on different roughened surfaces may differ greatly.

# 1. Introduction

Nucleate boiling heat transfer (NBHT) is one of the most favourable ways that provides large heat transfer coefficients, and it has been applied in numerous kinds of heat-exchange facilities. NBHT is also one of the most challenging problems in convective heat transfer. This is because from physical point of view phase change is a nano-scale phenomenon, and from engineering point of view the influencing factors are so many that until 90ties of last century we still did not know many things about NBHT even at phenomenological level (Lienhard, 1988).

It is now widely recognized that the cavities, cracks and crevices on the microstructure surfaces can easily become the nucleation sites because compared with a flat surface the energy required for forming a liquid-vapour surface is the minimum (Labuntzov, 1959). Besides, a reentrant cavity can trap the vapor and guarantee the nucleation continuously. According to this basic idea, extensive studies have been performed to enhance NBHT by changing the surface roughness (Ahn and Kim, 2011; Cooke and Kandlikar, 2011; Das et al., 2009; Dong et al., 2014; Dong et al., 2010; Honda and Wei, 2004; Kim, 2009; Shojaeian and Koşar, 2015). Followings are four representative papers. Das et al. (2009) developed several cavity structures, namely circular groove, rectangular groove and rounded base, to evaluate the performance of the cavities on the NBHT. The result suggested that the circular one performs the best. Honda and Wei (2004) studied several surface microstructures, such as drilled cavities and microfins, and proved the enhancement of NBHT and the increasing critical heat flux (CHF) with the microstructures. They found that the effects of microstructures on the enhancement of NBHT are mainly due to three aspects: the increase in heat transfer area, active nucleation site density and bubble departure frequency. Cooke and Kandlikar (2011) investigated the surface modifications to silicon chips, and demonstrated that the heat transfer coefficient can be improved by 3.4 times over that of a flat chip. Dong et al. (2014) conducted an experimental investigation of pool boiling with micro/nano-structures, and concluded that microstructures enhance bubble nucleation by increasing the nucleation site density, while the nano-structures increase the departure frequency.

Though plenty of efforts have been made on the structured surfaces, the details of boiling in such surfaces are still not understood completely. For example, we even cannot answer such a simple question as what kind of cavity has the best performance for NBHT. Therefore, our present knowledge of NBHT lacks physical basis for the designing a new microstructure surface.

With the development of computer technology, numerical simulations on NBHT have received much attention. Studies of direct numerical simulation on boiling were mostly reported in the past two decades where several kinds of interface tracking methods were used, such as VOF (Welch and Wilson, 2000), level-set (Mukherjee and Kandlikar, 2007) and VOSET method (Ling et al., 2014). However, with these methods, an initial nucleate site and a waiting period should be assigned which cannot describe the actual whole nucleation process.

In the recent decade, the lattice Boltzmann method (LBM) (He et al., 2009) has been carried out to simulate the two-phase flow with phase change. The existing models of boiling phase change may be classified into two groups (Dong et al., 2009; Gong and Cheng, 2012; Hazi and Markus, 2009; Shan and Chen, 1993). One is based on the phase-field method (Dong et al., 2009), which extends the Cahn-Hilliard equation by incorporating a phase-change source term into the system. The other is based on the pseudo-potential model (Shan and Chen, 1993). The energy equation is either solved with a thermal LB model (Gong and Cheng, 2012; Hazi and Markus, 2009) or other numerical methods (Li et al., 2015).

Márkus and Házi (2012b) investigated the effect of a cavity on the detachment of bubbles from a roughened surface with the pseudo-potential model. However, the simulations started with the assumption that the cavity was filled with vapor. Therefore, it could not be used to describe the actual whole nucleation process. Besides, the heat flux applied on the heating surface was non-uniform, which can be considered as an artificial input to promote the bubble nucleation.

In the present study, we used the Shan-Chen pseudo-potential model (Chen et al., 2014b; Shan and Chen, 1993) to simulate the multiphase flow, and the improved phase-change model proposed by Gong and Cheng (2012) to simulate the temperature transport. In our study no any assumptions of existing vapor embryo and no an artificial disturbance of temperature or density are needed. The rest of the paper is organized as follows. The Shan-Chen model and phase-change model are introduced in Sec.2. Results and discussion are given in Sec.3. Finally, some conclusions are made in Sec.4.

**Nomenclature**

| | | | |
|---|---|---|---|
| $c_p$ | specific heat capacity, J/(kg·K) | $t$ | time, s |
| $D_d$ | bubble diameter, m | $\rho$ | density, kg/m$^3$ |
| **e** | discrete velocity vector | $p$ | pressure, Pa |
| $f$ | bubble departure frequency, s$^{-1}$ | **u**, **u**$_r$ | velocity, real velocity |
| $f_i$ | density distribution functions | $\Delta\mathbf{x}, \Delta t$ | lattice space and time step |
| **F** | force, N | $\alpha$ | thermal diffusivity, m$^2$/s |
| $h_i$ | temperature distribution functions | $\lambda$ | thermal conductivity, W/(m·K) |
| **g** | gravity acceleration, m/s$^2$ | $v$ | viscosity, m$^2$/s |
| $q$ | local heat flux, W/m$^2$ | **Superscript** | |
| $Q$ | total heat flux, W/m$^2$ | $c$ | critical |
| $s$ | entropy | $f$ | fluid |
| $T$ | temperature, K | $s$ | solid |
| $T_s$ | saturation temperature, K | | |

## 2. Lattice Boltzmann method

### 2.1 The pseudo-potential LB model for multiphase flow

In the LB model, the motion of the fluid described by a set of density distribution functions is considered as a collective behavior of pseudo-particles on a mesoscopic level. Based on the simple and popular Bhatnagar-Gross-Krook (BGK) collision operator, the standard LB equation with a force term is described as follows

$$f_i(\mathbf{x}+c\mathbf{e}_i\Delta t, t+\Delta t) - f_i(\mathbf{x},t) = -\frac{1}{\tau_f}\left(f_i(\mathbf{x},t) - f_i^{eq}(\mathbf{x},t)\right) + F_i \quad (1)$$

where $f_i(\mathbf{x}, t)$ is the density distribution function at the lattice site **x** and time $t$, $\tau_f$ is the dimensionless relaxation time, $c=\Delta\mathbf{x}/\Delta t$ is the lattice speed with $\Delta\mathbf{x}$ and $\Delta t$ as the lattice spacing and time step, respectively. For a two-dimensional space with nine velocities at a given position D2Q9, the discrete velocities $\mathbf{e}_i$ are given by

$$\mathbf{e}_i = \begin{cases} 0 & i=0 \\ \left(\cos\left[\dfrac{(i-1)\pi}{2}\right], \sin\left[\dfrac{(i-1)\pi}{2}\right]\right) & i=1-4 \\ \sqrt{2}\left(\cos\left[\dfrac{(i-5)\pi}{2}+\dfrac{\pi}{4}\right], \sin\left[\dfrac{(i-5)\pi}{2}+\dfrac{\pi}{4}\right]\right) & i=5-8 \end{cases} \qquad (2)$$

The corresponding equilibrium distribution functions $f_i^{eq}(\mathbf{x},t)$ are given by

$$f_i^{eq}(\mathbf{x},t) = \omega_i \rho \left(1 + \dfrac{3}{c^2}(\mathbf{e}_i \cdot \mathbf{u}) + \dfrac{9}{2c^4}(\mathbf{e}_i \cdot \mathbf{u})^2 - \dfrac{3}{2c^2}\mathbf{u}^2\right) \qquad (3)$$

The weight factors $\omega_i$ are given by $\omega_0 = 4/9$, $\omega_{1-4} = 1/9$ and $\omega_{5-8} = 1/36$. The density $\rho$ and velocity $\mathbf{u}$ can be obtained from the first and second moments of the density distribution functions as follows:

$$\rho = \sum_i f_i(\mathbf{x},t) \qquad (4)$$

$$\rho \mathbf{u} = \sum_i \mathbf{e}_i f_i(\mathbf{x},t) \qquad (5)$$

The viscosity $v$ in the lattice unit is related to the relaxation time by

$$\upsilon = \dfrac{\Delta \mathbf{x}^2}{3\Delta t}(\tau_f - 0.5) \qquad (6)$$

We choose $\Delta \mathbf{x} = \Delta t = 1$ in the present study.

The force term $F_i$ is implemented with the exact difference method (Kupershtokh et al., 2009), which is given by

$$F_i = f_i^{eq}(\rho(\mathbf{x},t), \mathbf{u}+\Delta\mathbf{u}) - f_i^{eq}(\rho(\mathbf{x},t), \mathbf{u}) \qquad (7)$$

where $\Delta\mathbf{u} = \mathbf{F}\Delta t/\rho$, $\mathbf{F}$ is the total force acting on a fluid particle, which includes the fluid-fluid interaction force $\mathbf{F}_f$, fluid-solid interaction force $\mathbf{F}_s$ and other forces (such as the gravitational force $\mathbf{F}_g$). The interaction force is given by (Chen et al., 2014a; Chen et al., 2014b; Chen et al., 2013; Chen et al., 2015; Chen et al., 2014c; Kupershtokh et al., 2009)

$$\mathbf{F}_f = -G_f \psi(\mathbf{x}) c_s^2 \sum_{i=1}^{8} w(|\mathbf{e}_i|^2) \psi(\mathbf{x}+\mathbf{e}_i) \mathbf{e}_i \qquad (9)$$

where $G_f$ reflects the interaction strength among the fluid, and $w(|\mathbf{e}_i|^2)$ are the weights. In the present study, only the interactions of four nearest neighbors and four next-nearest neighbors are considered, and $w(1)=1/3$ and $w(2)=1/12$. The effective mass $\psi(\mathbf{x})$ is defined to calculate the interaction force based on local fluid density, and is expressed as (Chen et al., 2014b)

$$\psi(\mathbf{x}) = \sqrt{\frac{2}{G_f c_s^2}(p - \rho c_s^2)} \qquad (10)$$

A non-ideal fluid with the P-R equation of state (EOS) is considered (Yuan and Schaefer, 2006)

$$p = \frac{\rho RT}{1-b\rho} - \frac{a\alpha(T)\rho^2}{1+2b\rho-b^2\rho^2} \qquad (11)$$

$$\alpha(T) = \left[1 + (0.37464 + 1.54226\omega - 0.26992\omega^2) \times (1 - \sqrt{T/Tc})\right]^2 \qquad (12)$$

with $a = 0.45724(RT_c)^2/p_c$, $b = 0.1873RT_c/p_c$. Substituting Eqs. (11), (12) to Eq. (10), the effective mass can be calculated. Note that $G_f$ can be cancelled out and is put there only to ensure the right hand side of Eq. (10) to be reasonable. In order to simulate the nucleation process of water in the simulation, we choose $G_f = -1$, $a = 3/49$, $b = 2/21$, $R = 1$, $\rho_c=2.541858$, $T_c=0.109383$ and $\omega=0.344$. All the parameters in the present study are in lattice units, and the conversion between the lattice units and physical units can be found in (Yuan and Schaefer, 2006).

The fluid-solid force describing the wetting conditions is given by (Chen et al., 2014a;

Chen et al., 2014b; Chen et al., 2013; Chen et al., 2015; Chen et al., 2014c; Kupershtokh et al., 2009)

$$\mathbf{F}_s = -G_s \psi(\mathbf{x}) c_s^2 \sum_{i=1}^{8} w\left(|\mathbf{e}_i|^2\right) s(\mathbf{x}+\mathbf{e}_i) \mathbf{e}_i \tag{13}$$

where $s$ is an indicator function that equals 0 for fluid nodes and equals 1 for solid nodes. $G_s$ can be tuned to obtain different contact angles.

The gravitational force is expressed as

$$\mathbf{F}_g = (\rho - \rho_{average}) \mathbf{g} \tag{14}$$

with $\mathbf{g}$ is the gravitational acceleration and $\rho_{average}$ is the average fluid density of the entire domain. With Eq. (14), the mass average velocity of the system keeps constant since a zero external force averaged in the entire domain. According to (Chen et al., 2014b), $\mathbf{g}$ equals -0.0001 in the present study if without explicit explanation.

The physical velocity of the fluid is calculated as

$$\mathbf{u}_r = \mathbf{u} + \frac{\Delta t\, \mathbf{F}}{2\rho} \tag{15}$$

**2.2 The phase-change thermal LB model for energy equation**

Based on the entropy balance equation, Gong and Cheng (2012) proposed a source term for the energy equation, and with this improved energy equation, they have investigated the nucleate boiling (Gong and Cheng, 2013), laminar film condensation (Liu and Cheng, 2013) and boiling curve (Gong and Cheng, 2015). The entropy balance equation neglecting the viscous dissipation is given by

$$\rho T \frac{ds}{dt} = \nabla(\lambda \nabla T) \tag{16}$$

where $s$ is the entropy and $\lambda$ is the thermal conductivity. With the thermodynamic relation, the following equation can be derived

$$\frac{dT}{dt} = \frac{1}{\rho c_v}\nabla(\lambda \nabla T) - T\left(\frac{\partial p}{\partial T}\right)_\rho \nabla \cdot \mathbf{u}_r \qquad (17)$$

where $c_v$ is the specific heat at constant volume. With the assumption that $\rho c_v$ is constant in the entire domain, Equation (18) can be finally derived according to reference (Gong and Cheng, 2012).

$$\frac{\partial T}{\partial t} + \nabla \cdot (\mathbf{u}_r T) = \nabla(\alpha \nabla T) + T\left[1 - \frac{1}{\rho c_v}\left(\frac{\partial p}{\partial T}\right)_\rho\right]\nabla \cdot \mathbf{u}_r \qquad (18)$$

with $\alpha = \lambda/\rho c_v$. In order to solve Eq. (18), the thermal LB model is adopted

$$h_i(\mathbf{x} + c\mathbf{e}_i \Delta t, t + \Delta t) - h_i(\mathbf{x}, t) = -\frac{1}{\tau_T}(h_i(\mathbf{x}, t) - h_i^{eq}(\mathbf{x}, t)) + \omega_i \varphi \qquad (19)$$

where $h_i(\mathbf{x}, t)$ is the temperature distribution function, $\varphi$ is the source term in Eq. (18) and $\tau_T$ is the dimensionless relaxation time with $\tau_T = 0.5 + 3\alpha \Delta t/\Delta \mathbf{x}^2$. The equilibrium temperature distribution function is given by

$$h_i^{eq}(\mathbf{x}, t) = \omega_i T\left(1 + \frac{3}{c^2}(\mathbf{e}_i \cdot \mathbf{u}) + \frac{9}{2c^4}(\mathbf{e}_i \cdot \mathbf{u})^2 - \frac{3}{2c^2}\mathbf{u}^2\right) \qquad (20)$$

It is worth noting that the above described numerical methods are basically the same as those adopted (Li et al., 2015; Márkus and Házi, 2012a, b; Zhang and Chen, 2003). The treatments of nucleation formation include the small temperature fluctuations to the equation of state in the grid (Li et al., 2015), an initially assigned an existing embryo (Márkus and Házi, 2012a), and the fluctuation of density (Zhang and Chen, 2003) or non-uniform heat flux (Márkus and Házi, 2012b) applying on the wall. In these papers an artificial interruption should be added to start the nucleation; however, in our later presentation it will be shown that the above method can be used to predict the nucleation of a roughened surface without any artificial interruptions. In some sense the roughened surface can be regarded a kind of interruption, but it is the problem-inherent interruption, not artificially added, hence the present method can describe the actually whole nucleation process.

## 3. Validation of code and boundary conditions

### 3.1 Code validation

To validate our self-developed code, two representative problems are simulated. One is a bubble surrounded by liquid in a periodic zone, and the other is the comparison of the coexistence curves between the simulations and theoretical curves predicted by the Maxwell equal-area construction.

For the first problem, the Laplace law is tested by varying the bubble radius $R$. Initially, the temperature is set as $0.86T_c$, and the corresponding densities of liquid and vapor are 6.49895 and 0.37968, respectively. The variation of pressure drop $\Delta p$ with the $1/R$ is depicted in Figure. 1. It can be seen that the pressure drop varies with $1/R$ linearly. The slope, namely the surface tension coefficient, is 0.178.

For the second problem, a bubble is placed in a periodic domain filled with liquid. The density of the liquid is slightly larger than $\rho_c$ while the density of the vapor is slightly lower than $\rho_c$. By varying the saturation temperature $T$, the densities of the vapor and liquid would finally reach the corresponding value under the temperature $T$. Figure 2 shows the comparison of the coexistence curve obtained with the present model and the theoretical curves. The result shows that the prediction by the present model agrees well with the theoretical results.

### 3.2 Computational domain and boundary conditions

Simulations are performed on an open system filled with saturated water, and the water is heated through a roughened surface. Figure 3 depicts four different cavity grooves on the microstructured surface studied in this paper, namely the rectangular, circular, triangular and trapezoidal ones. The entire computational size is 61×601 (lattice unit), and the length $L$ and width of the cavity are 40 and 15, respectively. The thickness of solid $H_1$ and $H_2$ is 5 and 20, respectively. The fluid thermal diffusivity $\alpha$ is chosen to be equal to the viscosity $v$ in the present study. The thermal conductivity of the solid is assumed to be equal to that of the liquid for simplicity.

The detailed boundary conditions are as follows: periodic boundary condition is applied at the left and right side of the domain; convective boundary condition is defined at the outlet; a constant temperature $T_h = T_c$ is set on the bottom of the solid. Initially, the entire domain is filled with liquid water except for the solid. The density and the properties are determined with the corresponding saturation temperature $T_s = 0.86 T_c$.

## 4. Results and discussion

In the following section, result presentation will first be focused on the rectangular cavity. The effect of the cavity width on the bubble departure diameter and release frequency is investigated in detail. And then, the effect of different cavity grooves is studied. Finally, some results on different microstructured surfaces are presented.

### 4.1 Effect of rectangular cavity width

Bubble departure diameter $D_d$ and departure frequency $f$ are two critical characteristics during the nucleation process. Based on the static force balance between the buoyant and adhesive forces of a hanging drop on a horizontal surface, Fritz (1935) derived a relation between the departure diameter and gravity:

$$D_d \sim \left[\frac{\sigma}{g(\rho_l - \rho_g)}\right]^{0.5} \tag{21}$$

Figure 4 presents the simulation results of the effect of bubble departure diameter under various gravity accelerations. It is shown that the exponent of the fitting curve for our simulation results is -0.4875, which agrees quite well with the analytical result given by Fritz (1935).

The commonly used correlation of the bubble release frequency is described by Zuber (1963)

$$f^{-1} \sim D_b \left[\frac{\sigma g(\rho_l - \rho_g)}{\rho_l^2}\right]^{-0.25} \tag{22}$$

Combing with Eq. (21), one can find that the bubble release frequency correlates with the gravity acceleration as $g^{-0.75}$. Figure 5 shows the simulated results under various gravity accelerations. The exponent of the fitting curve is -1.029, which is about one third lower than the analytical result. This deviation may be explained as follows. The analytical result is derived based on a flat surface while our results are numerically obtained based on a surface with a cavity. For the surface with a cavity, the cavity is easily occupied by vapor after the departure of the former bubble, and the vapor can be considered as the nucleation site for the next vapor. Besides, the adhesive force is relatively smaller since the contact area between the vapor and the wall decreases, which means the buoyant force is predominant. Therefore, the bubble release frequency is much sensitive to the gravitation for a surface with cavity than that for a flat surface. Similar results are found in the numerical result of Márkus and Házi (2012b) and the experimental work of Sankaranarayanan et al. (2002) in which the exponent is around -0.827 and -1.05, respectively.

The nucleation process with different cavity widths is shown in Figure 6. Unlike most of the treatment for bubble formation, such as the artificial input of the disturbance of density or temperature (Li et al., 2015), non-uniform input heat flux (Márkus and Házi, 2012b), and a nucleate core set beforehand (Hazi and Markus, 2009), the generation of bubble in the present study is due to the geometry disturbance, which is much more reasonable as stated above. As can be seen in this figure, vapor is easily generated in a narrow cavity due to its high non-uniformity of temperature distribution of the liquid attached to the solid. The vapor is accumulated fast since the cavity space is relatively small. The generated bubble fills the whole cavity and spreads to the top surfaces of the solid (see $t =18000\text{-}24000\ \Delta t$) in the narrow cavity. Unlike the narrow cavity, the bubble in a wider cavity (cavity width higher than 10 lattices) generates and soon departs from the cavity without vapor accumulation process on the top surfaces. Therefore, the three-phase contact line of the wider cavity lies at the upper corners of the cavity while that of the narrow cavity lies away from the upper corners of the cavity.

Figure 7 shows the bubble departure snapshot under different cavity widths. Due to the interaction force of solid walls, the length of the vapor neck increases with the cavity width and then reaches some maximum value. Figure 8 shows the effect of cavity width on the bubble departure diameter and release frequency. As can be seen, the bubble departure diameter and the reverse of the bubble release frequency decrease with the cavity width, and then reach a constant.

The dimensionless total heat flux $Q'$ and dimensionless local heat flux $q'$ on the bottom of the solid are calculated as (Gong and Cheng, 2012)

$$T' = (T - T_s)/(T_h - T_s) \tag{22}$$

$$q'(x,t) = -\left(\frac{\partial T}{\partial y}\right)_{y=1} \frac{\Delta y}{T_c} \tag{23}$$

$$Q'(x,t) = \sum q'(x,t) \tag{24}$$

with $T'$ is the dimensionless temperature. Figure 9 depicts the dimensionless total heat flux under different cavity widths in one dimensionless period. It can be seen that the dimensionless total heat flux becomes more uniform with the increase in cavity width due to a more uniform temperature on the bottom of the cavity. And the value generally increases with the cavity width because narrower cavity tends to be choked with vapor. This result is agreeable well with the experimental study in Ghiu and Joshi (2005). The average dimensionless total heat fluxes in one period with the cavity width of 10, 14, 20, 24 and 30 lattices are 0.21853, 0.28682, 0.29569, 0.29711 and 0.29678 respectively. Figure 9 also shows that there are two peak values for the cavity width of 10 lattices during one period. One is at the moment of the bubble departure, and the other is at the moment when the three phase contact line lies at the cavity corners.

Figure 10 shows the distribution features of the dimensionless local heat flux on the bottom surface with different cavity widths. The dimensionless local heat flux distributions under the conditions when the dimensionless total heat flux depicted in Figure 9 is the highest and lowest are presented. It can be easily found that the distributions of local heat flux are relatively uniform with the increase in cavity width. In a narrow cavity, the local heat flux is relatively high at the center of the computational domain which explains that the vapor is easily formed. When the total heat flux is the lowest, the position of local peak value occurs under the cavity corners. However, when the total heat flux is the highest, the position of local peak value exist at the center of the computational domain with the cavity length under 20 lattices, and with the increase of the cavity length, the local value at the center of the domain tends to be smaller than that under the cavity corners, namely the three phase contact line.

**4.2 Effect of cavity grooves**

Figure 11 shows the bubble formation, growth and departure process in three different cavity grooves. When the departure of the first bubble is concerned, the bubble in the triangular groove departs at $t=15000\Delta t$, which is the fastest among all the grooves in the present study. This might be because that the triangular groove is not effective to trap more vapor. The vapor necks in other grooves are larger than that in triangular one.

When the first bubble detaches from the roughened surfaces, the remained vapor can be served as the nucleation site for the next bubble. It can also be found in Figure 11 and Figure 6 that a large amount of vapor is trapped in the circular, trapezoidal and rectangular grooves, while only a small amount of vapor exists in the triangular groove. Therefore, even though among all the grooves the first bubble is firstly released in the triangular groove, the bubble release frequency of the triangular cavity is relatively low. The detailed information of the departure diameter and release frequency for different cavity grooves is listed in Table 1. The table also shows that the release frequency of the circular groove is the highest. Therefore, it can be used to enhance the NBHT.

The density and temperature distributions at $t=30000\Delta t$ of four different cavity grooves are illustrated in Figure 12. One can see that the vapor temperature is lower than its surrounding because during the evaporation process heat flux goes from the surrounding to the bubble. In the cavity corners of the rectangular groove, the temperature is the lowest, which is also the three phase contact line. Figure 12 (b) also shows that the rectangular groove is the only structure which is fully occupied by the remained vapor.

The dimensionless total heat fluxes of different cavity grooves are shown in Figure 13. The figure shows that the circular groove performs the best for NBHT. This result agrees well with the experimental study (Das et al., 2009). The average dimensionless total heat fluxes during one period of the circular, trapezoidal, triangular and rectangular grooves are 0.53276, 0.48118, 0.26604 and 0.29678, respectively.

The detailed dimensionless local heat flux, when the dimensionless total heat flux is at its maximum and minimum value, is shown in Figure 14. The highest local heat flux for the circular, trapezoidal and triangular grooves lies at the middle part of the bottom, while the highest local heat flux for the rectangular groove lies under the two corners of the cavity. Generally, the local heat flux distributions for the rectangular and triangular grooves are much flatter compared with the other two grooves, this may be contributed to their specified structures.

**4.3 Effect of different roughened surfaces**

The effect of different roughened surfaces obtained by permutation and combination of these four cavity grooves on the nucleation boiling heat transfer are studied. Figure 15 shows the snapshot of the density distribution for different roughened surfaces. The merging of bubbles can be observed. This result indicates that even with the same microstructure cavities, the nucleate boiling process may be different.

The dimensionless total heat flux on the bottom of the solid wall is illustrated in Figure 16. One can find that the total heat fluxes are nearly the same for the four cases studied within the time of $15000\Delta t$, and then they differ from each other significantly (the inserted picture shows the magnified results). Combined with Figure 15, the bubbles are generated and grown independently ($0$-$15000\Delta t$), therefore, the total heat fluxes show the same tendency. With the proceeding of time, some bubbles are merged and the bubble morphologies vary greatly. This explains the variation of the total heat fluxes.

## 5. Conclusions

In this study, a phase-change lattice Boltzmann method is adopted and applied to the study of the nucleation boiling on surfaces with cavities. Without introducing any artificial interruptions, the actual nucleation process on roughened surfaces can be predicted by the adopted LBM. The main conclusions are as follows:

1. The bubble is easily generated and accumulated in a narrow cavity, and the dimensionless total heat flux generally increases with the cavity width.

2. The bubble departure diameter and the reverse of the bubble release frequency decrease with cavity width, and reach a stable value.

3. The surface with circular groove shows the best boiling heat transfer performance.

4. The bubble release frequency of the circular groove is the highest among all our studied grooves, and its local heat flux is also the highest in the most range of the width; the local heat flux distributions of the rectangular and triangular grooves are much flatter than the other two.

5. Different roughened surfaces obtained by permutation and combination of these four cavity grooves show different heat transfer performance.

**Acknowledgement**

This work is supported by the Key Project of the National Natural Science Foundation of China (51136004) and (51406145). We appreciate the helpful discussions with Doctor Kong Ling and Zhao-Hui Li from Xi'an Jiaotong University, China.

Table.1 Departure diameter and release frequency for different cavity grooves (lattice unit)

|       | Circular | Trapezoidal | Triangular | Rectangular |
|-------|----------|-------------|------------|-------------|
| $D_b$ | 41       | 39          | 42         | 37          |
| $f^{-1}$ | 5650  | 5715        | 7400       | 7785        |

# Figure Captions

Figure 1. Validation of the Laplace law.

Figure 2. Comparison of the coexistence curves.

Figure 3. Computational domain of different microstructure surfaces.

Figure 4. Bubble departure diameter under various $g$.

Figure 5. Bubble release frequency under various $g$.

Figure 6. Bubble formation, growth and departure process in cavities with width of 10, 14, 20, 24 and 30 lattices from left to right, respectively.

Figure 7. Bubble departure diameter of different cavities with width of 10, 14, 20, 24 and 30 lattices from left to right, respectively.

Figure 8. Effect of cavity width on the bubble departure diameter and release frequency.

Figure 9. Effect of cavity width on the dimensionless total heat flux.

Figure 10. Effect of the cavity width on the dimensionless local heat flux.

Figure 11. Bubble formation, growth and departure process under different cavity grooves.

Figure 12. Density and temperature distribution at t=30000 $\Delta t$, (a) temperature; (b) density.

Figure 13. Effect of cavity grooves on the dimensionless total heat flux.

Figure 14. Effect of cavity grooves on the dimensionless local heat flux.

Figure 15. Snapshot of the density distribution for different roughened structures, (a) Case 1; (b) Case 2; (3) Case 3; (4) Case 4.

Figure 16. Effect of roughened structures on the dimensionless total heat flux.

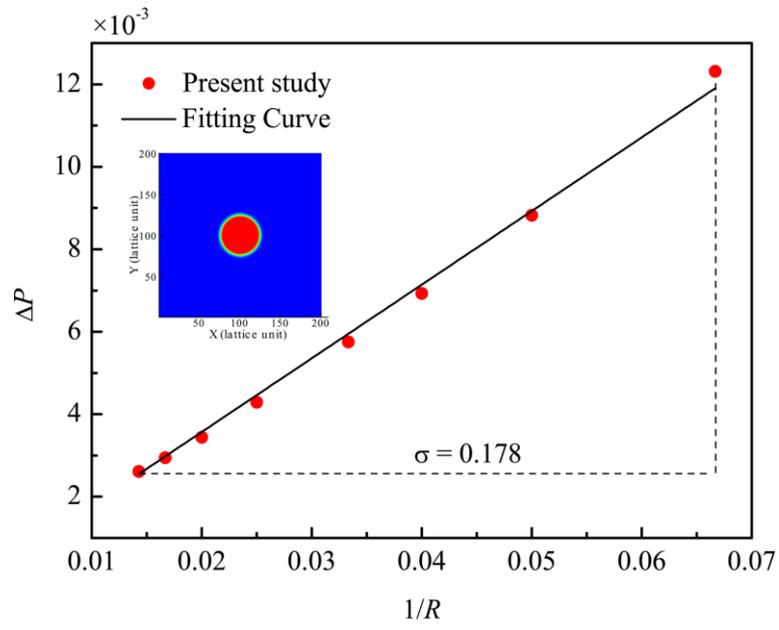

Fig. 1 Validation of the Laplace law

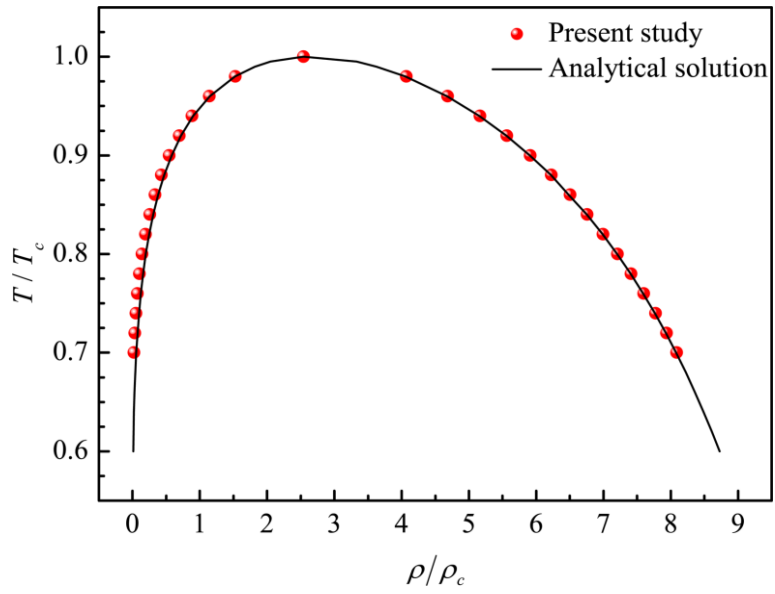

Fig. 2 Comparison of the coexistence curves

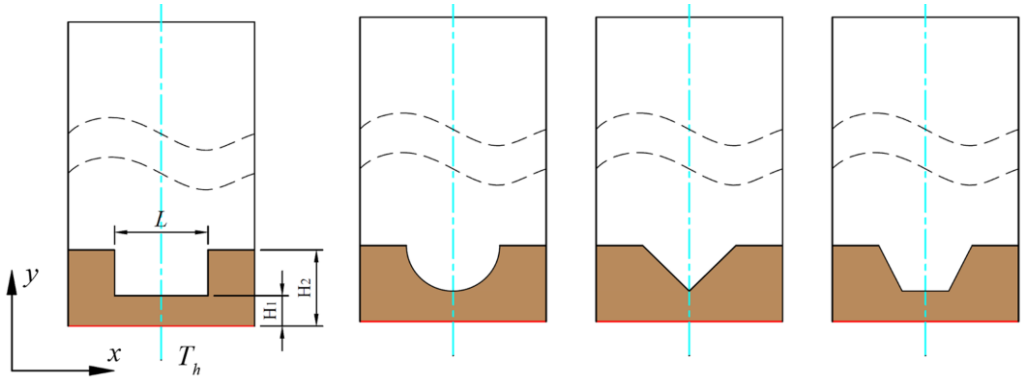

(a) Rectangular    (b) Circular  (c) Triangular  (d) Trapezoidal

Fig. 3 Computational domain of different microstructure surfaces

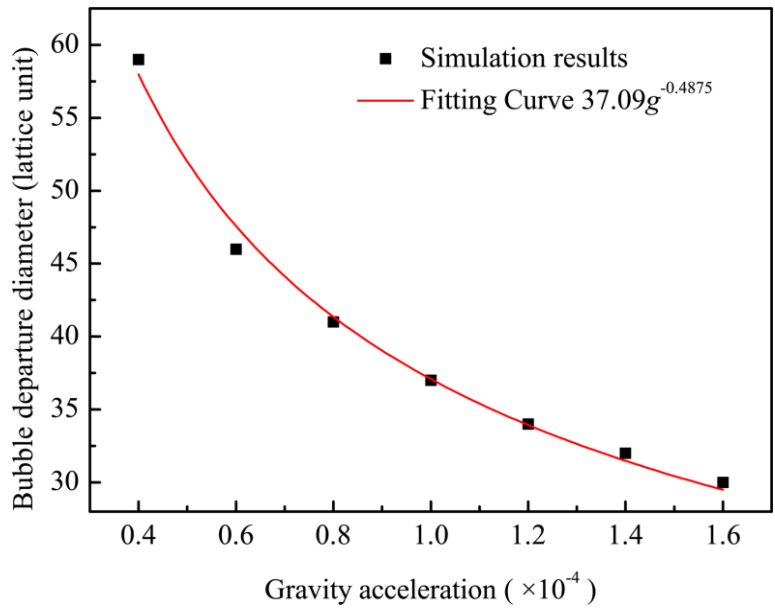

Fig. 4 Bubble departure diameter under various *g*

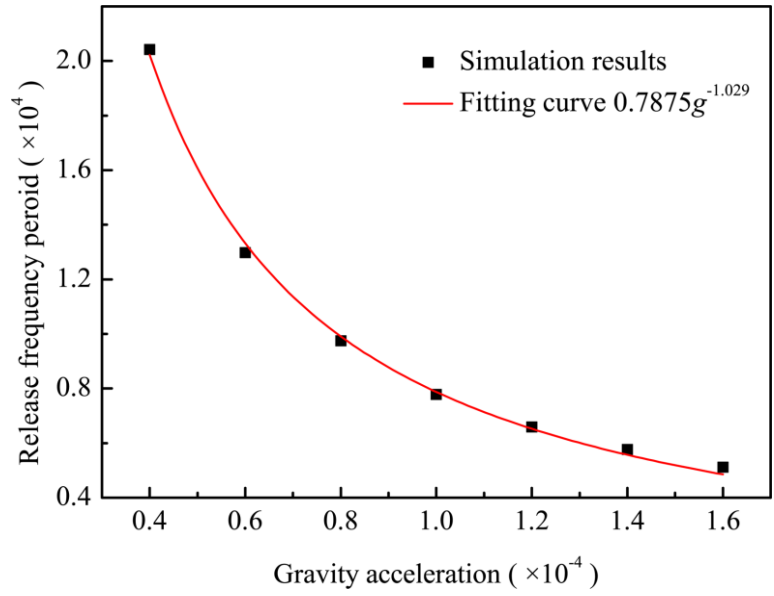

Fig. 5 Bubble release frequency under various *g*

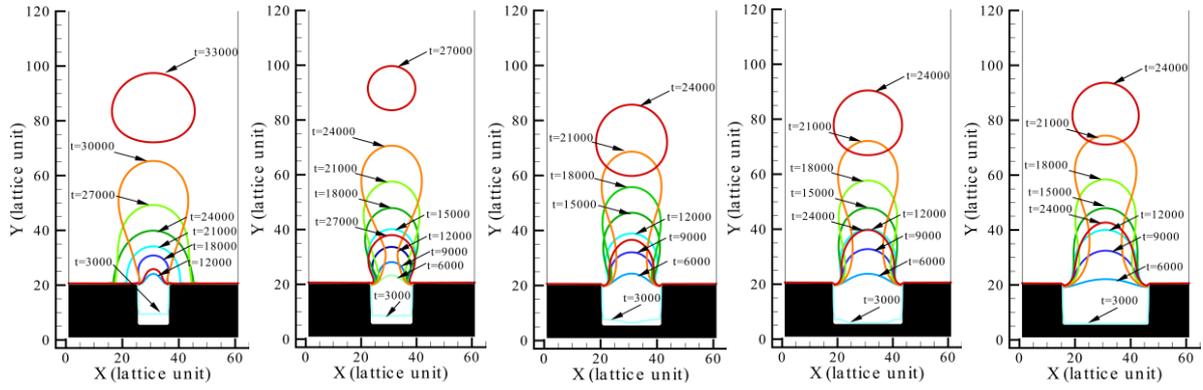

Fig. 6 Bubble formation, growth and departure process in cavities with width of 10, 14, 20, 24 and 30 lattices from left to right, respectively

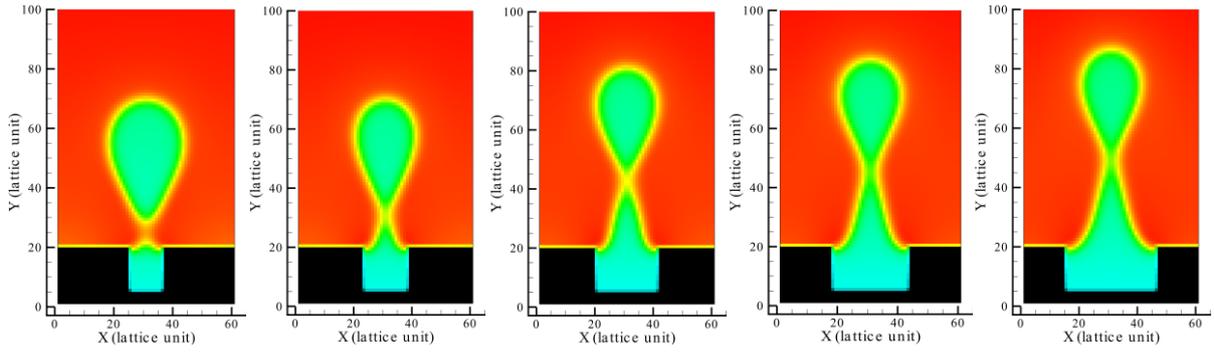

Fig. 7 Bubble departure diameter of different cavities with width of 10, 14, 20, 24 and 30 lattices from left to right, respectively

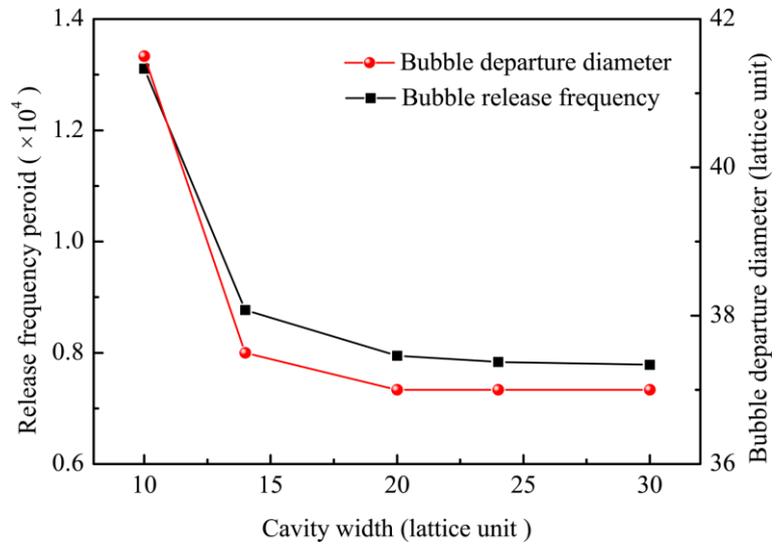
Fig. 8 Effect of cavity width on the bubble departure diameter and release frequency

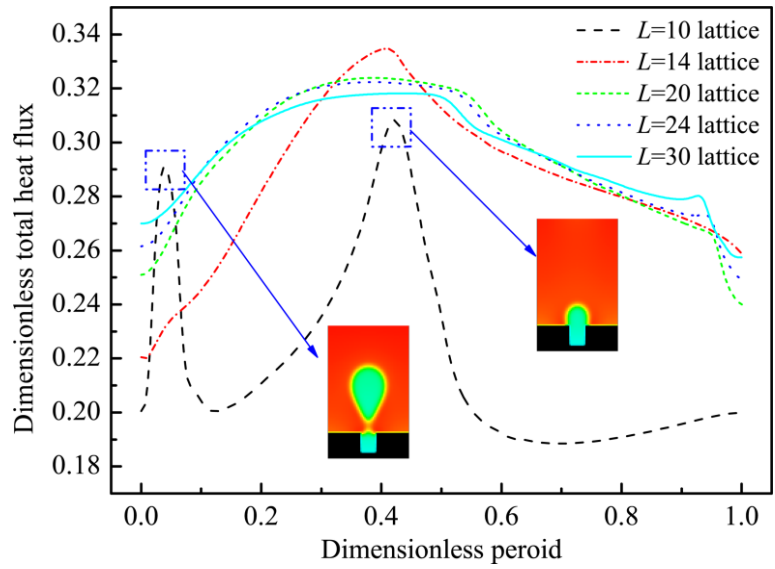

Fig. 9 Effect of the cavity width on the dimensionless total heat flux

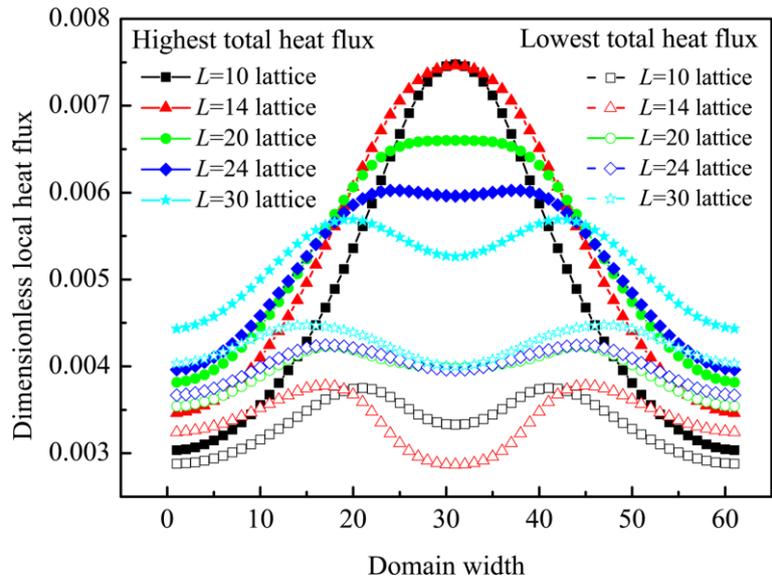

Fig. 10 Effect of the cavity width on the dimensionless local heat flux

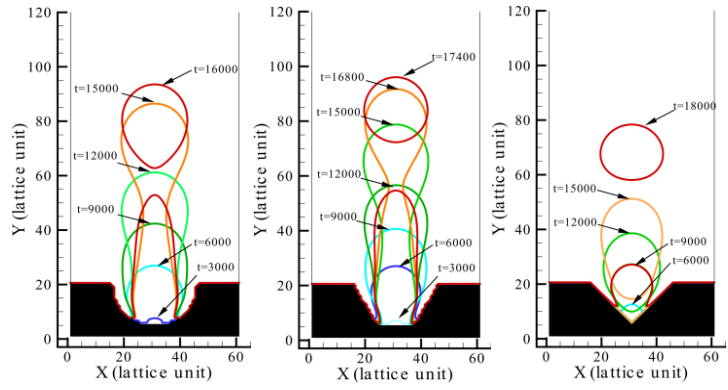

Fig. 11 Bubble formation, growth and departure process under different cavity grooves

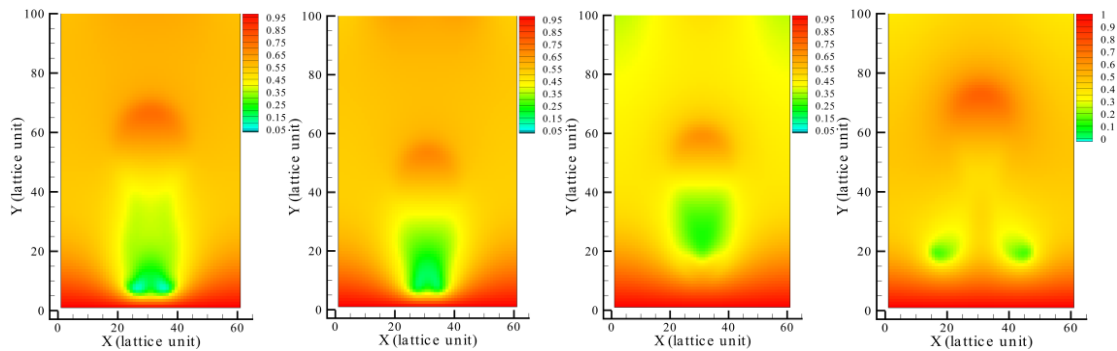

(a) Temperature

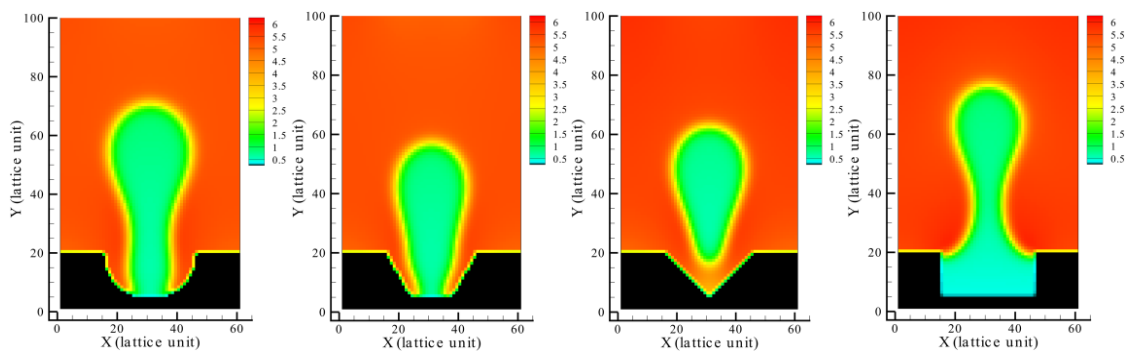

(b) Density

Fig. 12 Density and temperature distribution at t=30000 $\Delta t$.

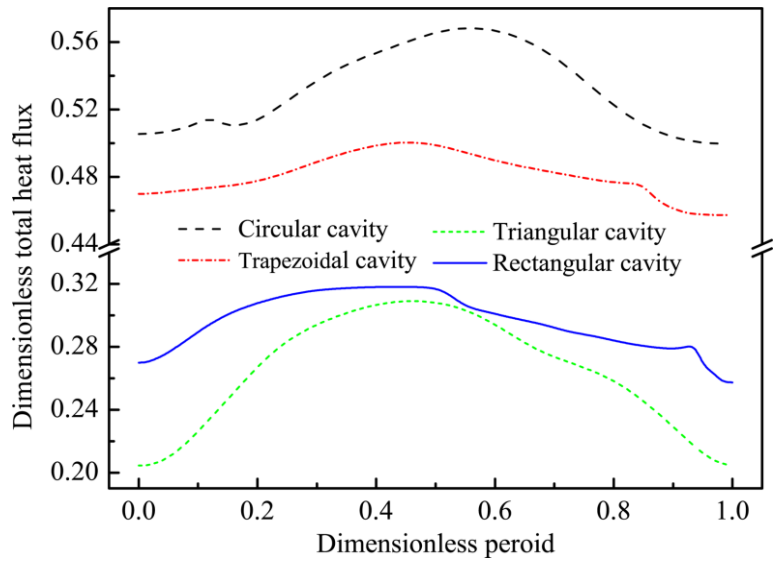

Fig. 13 Effect of cavity grooves on the dimensionless total heat flux

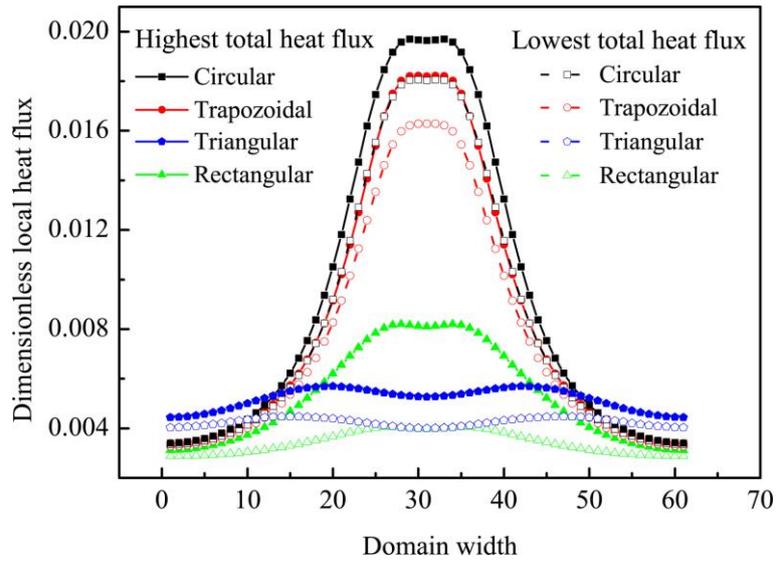

Fig. 14 Effect of cavity grooves on the dimensionless local heat flux

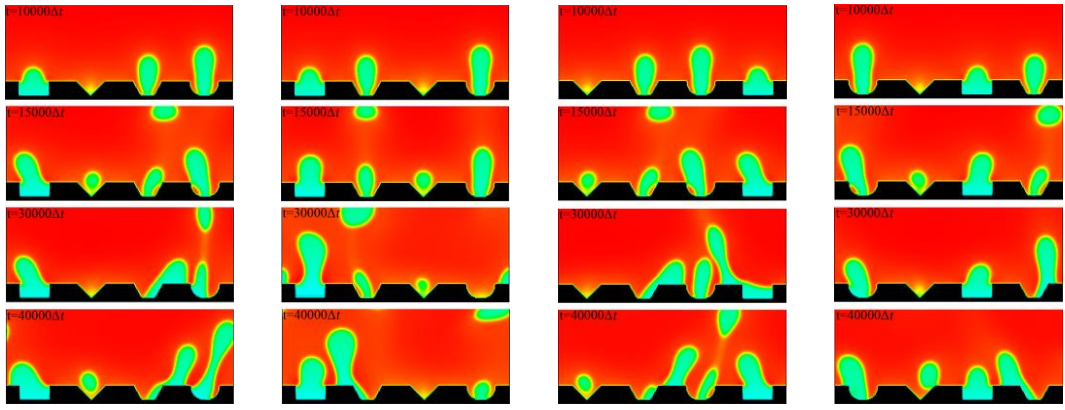

    (a) Case 1        (b) Case 2        (c) Case 3        (d) Case 4

Fig. 15 Snapshot of the density distribution for different roughened structures.

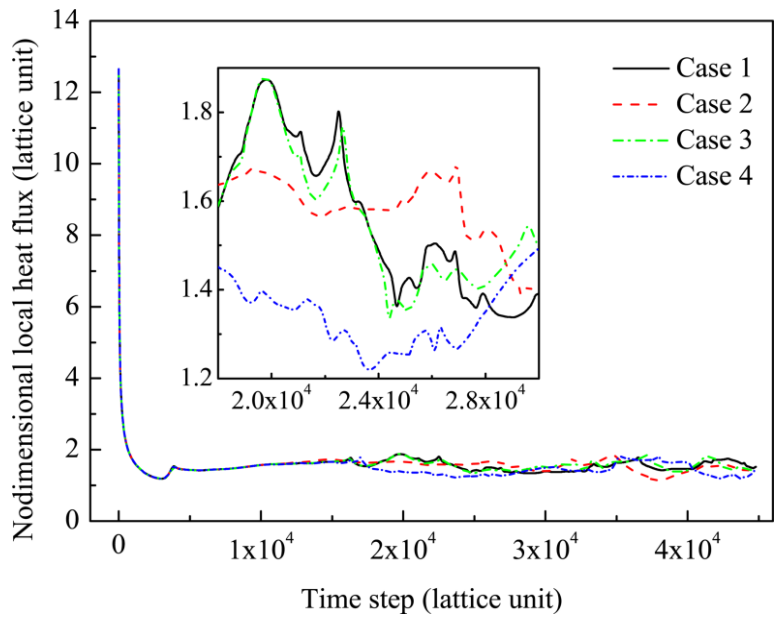

Fig. 16 Effect of roughened structures on the dimensionless total heat flux